\documentclass[aps,preprint,showpacs]{revtex4}
\usepackage{amssymb,graphicx}

\begin{document}

\title{String Cosmology of the D-brane Universe}

\author{Inyong Cho}
    \email{iycho@skku.edu}
    \affiliation{BK21 Physics Research Division and Institute of Basic Science,\\
        Sungkyunkwan University, Suwon 440-746, Korea}
\author{Eung Jin Chun}
    \email{ejchun@kias.re.kr}
    \affiliation{Korea Institute for Advanced Study,\\
        207-43 Cheongryangri 2-dong Dongdaemun-gu, Seoul 130-722, Korea}
\author{Hang Bae Kim}
    \email{hbkim@hanyang.ac.kr}
    \affiliation{BK21 Division of Advanced Research and Education in Physics,\\
        Hanyang University, Seoul 133-791, Korea}
\author{Yoonbai Kim}
    \email{yoonbai@skku.edu}
    \affiliation{BK21 Physics Research Division and Institute of Basic Science,\\
        Sungkyunkwan University, Suwon 440-746, Korea}

\begin{abstract}
We analyze homogeneous anisotropic cosmology driven by the dilaton and
the self-interacting ``massive'' antisymmetric tensor field
which are indispensable bosonic degrees with the graviton
in the NS-NS sector of string theories with D-branes.
We found the attractor solutions for this system, which show the overall
features of general solutions, and confirmed it through numerical analysis.
The dilaton possesses the potential due to the presence of the D-brane
and the curvature of extra dimensions.
In the presence of the non-vanishing antisymmetric tensor field,
the homogeneous universe expands anisotropically while the D-brane term
dominates. The isotropy is recovered as the dilaton rolls down
and the curvature term  dominates.
With the stabilizing potential for the dilaton, the isotropy
can also be recovered.
\end{abstract}


\maketitle

\section{Introduction}
\label{sec:Intro}

Low energy effective theories derived from the NS-NS sector of
string theories contain the gravity, $g_{\mu\nu}$, the dilation,
$\Phi$, and the antisymmetric tensor field, $B_{\mu\nu}$. The
existence of the last degree of freedom leads to  intriguing
implications in string cosmology \cite{Goldwirth:1993ha}.
In four spacetime dimensions,
the massless antisymmetric tensor field is dual to the a
pseudo-scalar (axion) field \cite{Kalb:1974yc}, and the
axion--dilaton system is known to develop an unobserved anisotropy
in our Universe, which can be diluted away at late times only in a
contracting universe \cite{Copeland:1994km}.

Such a disastrous cosmological situation can be resolved also in
string theory which contains another dynamical object called D-brane
\cite{Witten:1995im}. The gauge invariance on the D-brane is
maintained through the coupling of the gauge field strength to the
antisymmetric tensor field \cite{Cremmer:1973mg,Witten:1995im,Pol}.
Then, the effective action derived on the D-brane describes the
antisymmetric tensor field as a massive and self-interacting 2-form
field. Cosmological evolution of such a tensor field has been
investigated in Ref.~\cite{Chun:2005ee} assuming that the dilaton is
fixed to a reasonable value. Although the time-dependent magnetic $B$
field existing in the early universe develops an anisotropy in the
universe, it was realized that the matter-like behavior of the $B$
field (B-matter) ensures a dilution of the anisotropy at late times
and thus the isotropy is recovered in reasonable cosmological
scenarios \cite{Chun:2005ee}. In such sense the effect of antisymmetric
tensor field on the D-brane is distinguished from that of field strength
of the U(1) gauge field \cite{Kim:2004zq}.

In this paper, we investigate the cosmological evolution of the
B-matter--dilaton system in our Universe which is assumed to be
imbedded in the D-brane. The usual string cosmology with the
dilaton suffers from the notorious runaway problem, which is also
troublesome in the D-brane universe.  In our study, the dilaton
obtains two exponential potential terms due to the curvature of
extra dimensions $\Lambda$, and the D-brane tension
(the mass term of the B-matter) $m_B$.
It is interesting to
observe that  the dilaton can be stabilized for negative $\Lambda$
\cite{Kim:2005kr} which,  however,  leads to a contracting
universe due to the effective negative cosmological constant in
our Universe.  When $\Lambda$ is positive, the B-matter dominance
will be overturned by the $\Lambda$ dominance as the dilaton runs
away to the negative infinity. As a consequence of this,  the
initial anisotropy driven by the B-matter can also be diluted away
at late times.

For a realistic low energy effective theory,  string theory must
be endowed with a certain mechanism generating an appropriate
vacuum expectation value for the dilaton.  In such a situation,
the dilaton is expected to be stabilized at some stage of the
cosmological evolution affecting the dynamics of the antisymmetric
tensor field. Taking an example of the dilaton stabilization, we
will also examine  the cosmological evolution of the B-matter and
the dilaton in which the essential features of
Ref.~\cite{Chun:2005ee} are reproduced.

This paper is organized as follows.  In Section~\ref{sec:2}, we
describe the low energy effective action of the D-brane universe
and the corresponding field equations.  Before our main
discussion, Section~\ref{sec:3} is devoted to presenting homogeneous
solutions in flat spacetime illustrating some interesting features
of the B-matter--dilaton system in a simple way.  In Section~\ref{sec:4}, we
find semi-analytic and numerical cosmological solutions to observe
an intriguing interplay of the curvature $\Lambda$ and the D-brane
tension $m_B$. In Section~\ref{sec:5}, we consider the evolution of the
anisotropic universe with the dilaton stabilization which leads to
a satisfactory cosmology of the D-brane universe. We conclude in
Section~\ref{sec:Conc}.

\section{String Effective Theory on the D-brane }
\label{sec:2}

The main idea of the D-brane world is that we reside on a D$p$-brane
imbedded in 10 (or 11) dimensional spacetime with extra-dimensions
compactified. The bosonic NS-NS sector of the D-brane world consists of
the U(1) gauge field $A_\mu$ living on the D$p$-brane and the bulk
degrees including the graviton $g_{\mu\nu}$, the dilaton $\Phi$, and
the antisymmetric tensor field of rank-two $B_{\mu\nu}$. In the
presence of the brane, the gauge invariance of $B_{\mu\nu}$ is restored
through its coupling to a U(1) gauge field $A_\mu$ and the gauge
invariant field strength is \cite{Witten:1995im}
\begin{equation}
{\cal B}_{\mu\nu} \equiv B_{\mu\nu}+2\pi\alpha'F_{\mu\nu},
\qquad{\rm where}\qquad F_{\mu\nu}=\partial_\mu A_\nu-\partial_\nu
A_\mu.
\end{equation}

Even though we are assumed to live on the D-brane,
we adopt the conventional compactification in the sense that
the extra dimensions are compact and stabilized, and thus static.
In the presence of the D-brane, this would in general require
an additional setup like additional branes and fluxes along the extra
dimensions~\cite{Giddings:2001yu}, and the working of non-perturbative
effects~\cite{Kachru:2003aw}.
The extra dimensions are warped due to branes and fluxes.
The warping of extra dimensions gives a warp factor in the definition
of the four-dimensional Planck scale~\cite{Randall:1999ee},
and also induces the potential for the dilaton~\cite{Giddings:2001yu}
as described below.
If the fluxes in the extra dimensions significantly affect
the compactification,
the existence of the fluxes on the D-brane seems also natural and
it is intriguing to tackle their effect in the early universe.
However, simultaneous consideration of the fluxes along both the extra
dimensions and the D-brane lets the computation almost intractable.
Therefore, the conventional compactification is an appropriate setup
at the present stage, and then we take the following
four-dimensional effective action of the bosonic sector
in the string frame~\cite{Chun:2005ee}
\begin{eqnarray}
\label{action-S} S_{{\rm S}}&=& \frac{1}{2\kappa_4^2} \int d^4\tilde
x \sqrt{-\tilde g} \left[ e^{-2\Phi}\left(\tilde
R-2\Lambda+4\tilde\nabla_\mu\Phi\tilde\nabla^\mu\Phi -\frac{1}{12}
\tilde H_{\mu\nu\rho}\tilde H^{\mu\nu\rho}\right)
\vphantom{\sqrt{\frac12}}\right. \nonumber\\ && \hspace{25mm} \left.
-{m_B^2}e^{-\Phi} \sqrt{1 + \frac12\tilde{\cal
B}_{\mu\nu}\tilde{\cal B}^{\mu\nu}
- \frac{1}{16}\left(\tilde{\cal B}^*_{\mu\nu}\tilde{\cal B}^{\mu\nu}\right)^2}
\ \right],
\end{eqnarray}
where the tilde denotes the string frame quantity,
$H_{\mu\nu\rho}=\partial_{[\mu}{\cal B}_{\nu\rho]}$ and
${\cal B}^{\ast}_{\mu\nu}=\frac12\sqrt{-g}\epsilon_{\mu\nu\alpha\beta}
{\cal B}^{\alpha\beta}$ with $\epsilon_{0123} = 1$.
The parameter $m_B$ is defined by $m_B^2=2\kappa_4^2{\cal T}_p$
where ${\cal T}_p$ is the effective brane tension. Note that we omitted
the Chern-Simons like term assuming the trivial R-R background.
If we assume that the six
extra-dimensions are compactified with a common radius $R_{{\rm c}}$,
%and the dilaton is stabilized to give a finite string coupling
%$g_{{\rm s}}=\langle e^\Phi\rangle$,
one finds $m_B=\pi^{\frac14}(g_{{\rm s}}^2/4\pi)^{\frac{p-3}{16}}
\left(R_{{\rm c}}M_{{\rm P}}\right)^{\frac{15-p}{8}}M_{{\rm P}}$
where $g_s=e^\Phi$ and $M_{{\rm P}}=2.4\times10^{18}{\rm GeV}$ is
the four-dimensional Planck mass.
The qualitative features of our results do not depend on specific
values of $p$.  Thus, we take $p=3$ for simplicity.
The $\Lambda$ term comes from the scalar curvature
or the condensate $\langle H^2 \rangle$ of extra
dimensions integrated over the whole extra dimensions.

The action, and thus also field equations, can be written in a more
familiar form in the Einstein metric, which is defined by
\begin{equation}
g_{\mu\nu} = e^{-2\Phi}\tilde g_{\mu\nu}.
\end{equation}
We will work in this metric from now on. In terms of Einstein
metric, the action becomes
\begin{eqnarray}
\label{action-E}
S_{\rm E}&=& \frac{1}{2\kappa_4^2} \int d^4x
\sqrt{-g} \left[ R -2\Lambda e^{2\Phi} -2(\nabla\Phi)^2
-\frac{1}{12}e^{-4\Phi}H^2 \vphantom{\sqrt{\frac12}}\right.
\nonumber\\ && \hspace{25mm} \left. -{m_B^2}e^{3\Phi}\sqrt{1 +
\frac12e^{-4\Phi}{\cal B}^2 - \frac{1}{16}e^{-8\Phi}\left({\cal
B}^*{\cal B}\right)^2} \right].
\end{eqnarray}
The field equations derived from the action (\ref{action-E}) are
\begin{equation}
\label{B-eq1} \nabla^\lambda H_{\lambda\mu\nu}
-4H_{\lambda\mu\nu}\nabla^\lambda\Phi -m_B^2e^{3\Phi}\frac{{\cal
B}_{\mu\nu}-\frac14e^{-4\Phi}{\cal B}^*_{\mu\nu} \left({\cal B}{\cal
B}^*\right)}{\sqrt{1+\frac12e^{-4\Phi}{\cal B}^2
-\frac{1}{16}e^{-8\Phi}\left({\cal B}{\cal B}^*\right)^2}} = 0,
\end{equation}
\begin{equation}
\label{P-eq1} -\nabla^2\Phi + \frac{\partial V(\Phi)}{\partial\Phi}
= 0,
\end{equation}
\begin{equation}
\label{E-eq1} G_{\mu\nu} = \kappa_4^2 T_{\mu\nu},
\end{equation}
where the dilaton potential is
\begin{eqnarray}
V(\Phi) &=& \frac14\left[2\Lambda e^{2\Phi} +
\frac{1}{12}e^{-4\Phi}H^2
\vphantom{\sqrt{\frac12}}\right.\nonumber\\&&\hspace{16mm}\left.
{}+m_B^2e^{3\Phi}\sqrt{1 + \frac12e^{-4\Phi}{\cal B}^2 -
\frac{1}{16}e^{-8\Phi}\left({\cal B}^*{\cal B}\right)^2} \ \right],
\label{dilaton-potential}
\end{eqnarray}
and the energy-momentum tensor is given by
\begin{eqnarray}
\kappa_4^2 T_{\mu\nu}&=& -g_{\mu\nu}\Lambda e^{2\Phi}
+2\nabla_\mu\Phi\nabla_\nu\Phi-g_{\mu\nu}(\nabla\Phi)^2 \nonumber\\
&& {}+\frac{1}{12}e^{-4\Phi}
\left(3H_{\mu\lambda\rho}H_\nu^{\;\lambda\rho}
-\frac12g_{\mu\nu}H^2\right) \nonumber\\ && {}+\frac12m_B^2e^{3\Phi}
\frac{-g_{\mu\nu}-\frac12g_{\mu\nu}e^{-4\Phi}{\cal B}^2
+e^{-8\Phi}{\cal B}_{\mu\lambda}{\cal B}_\nu^{\;\lambda}}
{\sqrt{1+\frac12e^{-4\Phi}{\cal B}^2
-\frac{1}{16}e^{-8\Phi}\left({\cal B}{\cal B}^*\right)^2}}.
\label{TB}
\end{eqnarray}

In the subsequent sections, we examine the equations of motion
(\ref{B-eq1})--(\ref{E-eq1}) and find homogeneous solutions.
Cosmological implication of the D-brane universe is of our main
interest, including stabilization of the dilaton and condensation of
the antisymmetric tensor field.

\section{Homogeneous Solutions in Flat Spacetime}
\label{sec:3}

For the sake of the simplicity, let us first look into the
dynamics of our system in flat spacetime. With
$g_{\mu\nu}=\eta_{\mu\nu}$, the Einstein equations (\ref{E-eq1})
become simple and gauge invariance of the field variables ${\bf
E}^{i}\equiv {\cal B}_{i0}$ and ${\bf B}^{i}\equiv
\epsilon_{0ijk}{\cal B}_{jk}/2$ which directly appear in the
expression of energy-momentum (\ref{TB}) allows a homogeneous
configuration
\begin{equation}\label{hom}
\Phi=\Phi(t),\quad {\bf E}={\bf E}(t),\quad {\bf B}={\bf B}(t).
\end{equation}

For homogeneous configuration (\ref{hom}), $0i$-components ({\it
electric} components) of the equation of the antisymmetric tensor
field (\ref{B-eq1}) reduce to an algebraic equation
\begin{equation}\label{0i}
m_{B}^{2}e^{3\Phi} \frac{{\bf E}+e^{-4\Phi}{\bf B}({\bf E}\cdot{\bf
B})}{ \sqrt{1-e^{-4\Phi}({\bf E}^{2}-{\bf B}^{2}) -e^{-8\Phi}({\bf
E}\cdot {\bf B})^{2}}} =0.
\end{equation}
$ij$-components ({\it magnetic} components) give
\begin{equation}\label{ij}
\ddot{\bf B}-4\dot{\bf B}\dot\Phi = -e^{3\Phi} \frac{{\bf
B}-e^{-4\Phi}{\bf E} ({\bf E}\cdot{\bf B})}{\sqrt{1-e^{-4\Phi}({\bf
E}^{2}-{\bf B}^{2}) -e^{-8\Phi}({\bf E}\cdot {\bf B})^{2}}},
\end{equation}
where the overdot denotes differentiation with respect to
%$\partial/\partial{\tilde t}$ over
the rescaled dimensionless time
\begin{equation}\label{ti}
{\tilde t}=m_{B}t.
\end{equation}
These equations (\ref{0i})--(\ref{ij}) tell us that the {\it
magnetic} components ${\bf B}$ are {\it dynamical} but the {\it
electric} components are determined by the {\it constraint}
equation (\ref{0i}). Dynamics of the dilaton is governed by
(\ref{P-eq1}) which becomes
\begin{equation}\label{de}
\ddot\Phi+ \frac{1}{4}e^{-4\Phi}\dot{\bf B}^{2}= -
\lambda e^{2\Phi}-\frac{1}{4}e^{3\Phi} \frac{3-e^{-4\Phi}({\bf
E}^{2}-{\bf B}^{2}) +e^{-8\Phi}({\bf E}\cdot {\bf
B})^{2}}{\sqrt{1-e^{-4\Phi} ({\bf E}^{2}-{\bf B}^{2})
-e^{-8\Phi}({\bf E}\cdot {\bf B})^{2}}}\, ,
\end{equation}
where the dimensionless parameter $\lambda$ is defined by
\begin{equation}\label{co}
\lambda=\frac{\Lambda}{m_{B}^{2}}.
\end{equation}

The constraint equation (\ref{0i}) forces its numerator to vanish
except a trivial solution $\Phi=-\infty$, however vanishing
numerator allows only a {\em vanishing} electric field solution
${\bf E}=0$. Then the equations for the magnetic components
(\ref{ij}) and the dilaton (\ref{de}) reduce to
\begin{eqnarray}
\ddot{\bf B}-4\dot{\bf B}\dot\Phi &=& -e^{3\Phi} \frac{{\bf
B}}{\sqrt{1+e^{-4\Phi}{\bf B}^{2}}},
\label{m1}\\
\ddot\Phi+ \frac{1}{2}e^{-4\Phi}\dot{\bf B}^{2}&=& -
\lambda e^{2\Phi} -\frac{1}{4}e^{3\Phi}\frac{3+e^{-4\Phi}{\bf
B}^{2}}{ \sqrt{1+e^{-4\Phi}{\bf B}^{2}}}\, . \label{m2}
\end{eqnarray}
In the following, we will find solutions of (\ref{m1}) and (\ref{m2})
to see how the dilaton and the magnetic component behave.

\subsection{Dilaton potential with ${\bf B}=0$}
\label{subsec:3.A}

\begin{figure}[ht]
\begin{center}
\includegraphics{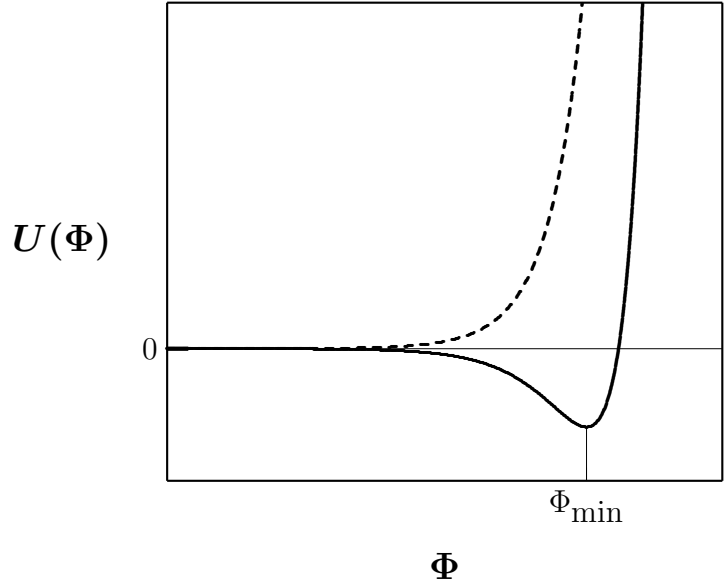}
\end{center}
\caption{\small The graph of $U(\Phi)$. For
$\lambda\ge 0$, $U(\Phi)$ monotonically increases (dashed
line). For $\lambda<0$, $U(\Phi)$ has a minimum at
$\Phi=\ln(-4\lambda/3)$ (solid line). }
\label{fig1}
\end{figure}

When the magnetic components of the antisymmetric tensor field ${\bf
B}$ vanish, we have only the dilaton equation from (\ref{m2})
\begin{equation}\label{d1}
\ddot\Phi=-\frac{dU}{d\Phi},
\end{equation}
where the dilaton potential $U(\Phi)$ is given by the sum of the
curvature term and the brane tension term
\begin{equation}\label{pu}
U(\Phi)=\frac{\lambda}{2}e^{2\Phi}
+\frac{1}{4}e^{3\Phi}
\end{equation}
as shown in Figure~\ref{fig1}.

For $\lambda\ge 0$, the dilaton potential $U(\Phi)$ monotonically increases,
so the dilaton finally rolls to the minimum $\Phi(t)\rightarrow -\infty$ as
$t\rightarrow\infty$.  For $\lambda<0$, the dilaton potential
$U(\Phi)$ has the minimum value $U_{{\rm min}}= 8{\lambda}^{3}/27$
at $\Phi_{{\rm min}} =\ln(-4\lambda/3)$. If $E=U_{{\rm min}}$, the
dilaton is stuck at $\Phi=\Phi_{{\rm min}}$. When $U_{{\rm
min}}<E<0$, the dilaton oscillates around the minimum. Therefore,
the dilaton is stabilized with mass $m_{\Phi}=\frac43\lambda^{3/2}m_B$,
and due to exponential potential terms,
its mass can be much different from the vacuum expectation
value of the dilaton $m_{B}\Phi_{{\rm min}}$. As $\lambda$ increases,
the dilaton mass rapidly grows.  When $E\ge
0$, the dilaton again rolls to negative infinity. It means that
the dilaton is easily destabilized by its fluctuations, if the
dilaton mass becomes too small. Though the negative vacuum energy
$U_{{\rm min}}<0$ does not affect the dynamics of the dilaton and
the antisymmetric tensor field in flat spacetime, its coupling to
gravity leads to singular cosmological evolution as we shall see
in Section~\ref{sec:4}. Therefore, for the proper description of cosmology,
some additional term is needed to adjust the value of $U_{{\rm
min}}$ to vanish as we shall discuss in Section~\ref{sec:5}.

The runaway or stabilized behavior of the dilaton remains
essentially unchanged even with $B\neq0$  although the dependence
on the initial conditions alters as we will see in the next subsection.

\subsection{Magnetic solution}
\label{subsec:3.B}

Without loss of generality,  let us assume that the direction of
the magnetic component of antisymmetric tensor field is fixed as
${\bf B}(t)=B(t){\bf k}$. Although the equations of motion
(\ref{m1})--(\ref{m2}) are still complicated even under this
assumption, the presence of the magnetic components do not alter the
story of the dilaton much.

First, let us consider a simple case of zero brane tension
($m_B=0$), which is analytically tractable.
Then in (\ref{m1})--(\ref{m2}), only the $\lambda$-dependent term survives
on the right-hand side. (For this case, we will use the original $\Lambda$ and time $t$
since $m_B=0$.)
The solutions to these equations are
\begin{equation}
\Phi (T) = T - {1\over 2}\ln\left[ \left( {1\over 4}e^{2T} +\Lambda\right)^2
+b_1^2c_1^2\right] +\ln c_1 \,,
\end{equation}
and
\begin{equation}
\pm B(T) = 16{c_1\over b_1} {\Lambda e^{2T} +4(\Lambda^2+b_1^2c_1^2) \over
e^{4T} +8\Lambda e^{2T} +16(\Lambda^2+b_1^2c_1^2)}
+4{\Lambda\over b_1^2} \tan^{-1}\left({e^{2T}+4\Lambda \over 4b_1c_1}\right)
 +b_2\,,
\label{BBB}
\end{equation}
where $b_1$, $b_2$, and $c_1$ are integration constants, and
the time has been rescaled to $T=c_1(t+c_2)$ by absorbing another
constant $c_2$.
The solutions are still remaining valid under $T \to -T$,
i.e., under the change of the sign of $c_1$.
Therefore, we can fix the time $T$ to flow to the positive direction and
$c_1 >0$.
The solutions approach their asymptotic configurations at large $T$,
\begin{eqnarray}
\Phi (T) &=& -T +\ln (4c_1)\,,\\
\pm B(T) &=& \mbox{sign}(b_1) {2\pi \Lambda \over b_1^2} +b_2 \equiv
B_{{\rm st}}\,.
\end{eqnarray}
We observe that the dilaton rolls linearly in time to negative infinity,
and the antisymmetric tensor field condensates to a constant $\pm B_{{\rm st}}$
which depends on $\Lambda$.
There are two topologically distinct condensation processes depending
on the signature of $b_1$ which is to be set by the initial conditions;
the change in $\Lambda$ contributes to $B_{{\rm st}}$ oppositely.

With the brane tension term turned on,
we numerically study due to the complexity of field equations.
The solutions are classified into two classes as in the pure dilaton
case in Section~\ref{subsec:3.A},
depending on the signature of $\lambda$.

\begin{figure}[ht]
\begin{center}
\includegraphics[height=60mm]{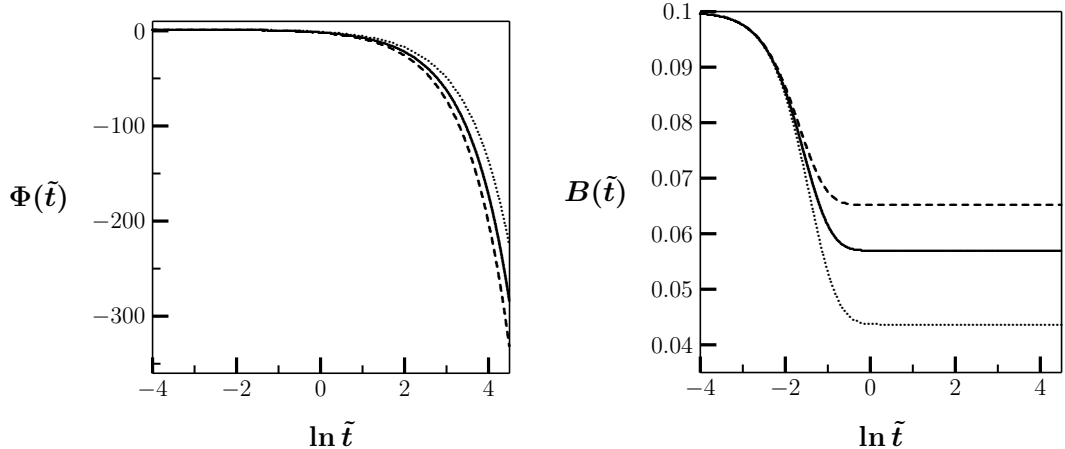}
\end{center}
\caption{\small $\Phi$ and $B$ in the Einstein frame for $
\lambda =0$ (solid lines), $\lambda = 1/2$
(dashed lines), and $\lambda = -1/2$ (dotted lines).
The initial conditions are $\Phi_0=1$ and $B_0=0.1$. The
configurations are very similar regardless of $
\lambda$.}
\label{fig2}
\end{figure}

For $\lambda\ge 0$, the dilaton approaches negative infinity
irrespective of initial conditions, $\Phi_0$ and $B_0$. The
condensed value of the magnetic component deviates from the
initial value in the beginning, and
then stabilizes to a constant as shown in Figure~\ref{fig2}.
We observed from the numerical results that there exist two branches
of the magnetic condensation depending on the initial conditions
as discussed above,
but we show only one branch in the figure.
Like in the $m_B=0$ case, once the magnetic component is condensed,
the condensation survives as a constant value. It is different from
the case without the dilaton, in which the magnetic component
permanently oscillates in flat spacetime~\cite{Chun:2005ee}.

\begin{figure}[ht]
\includegraphics{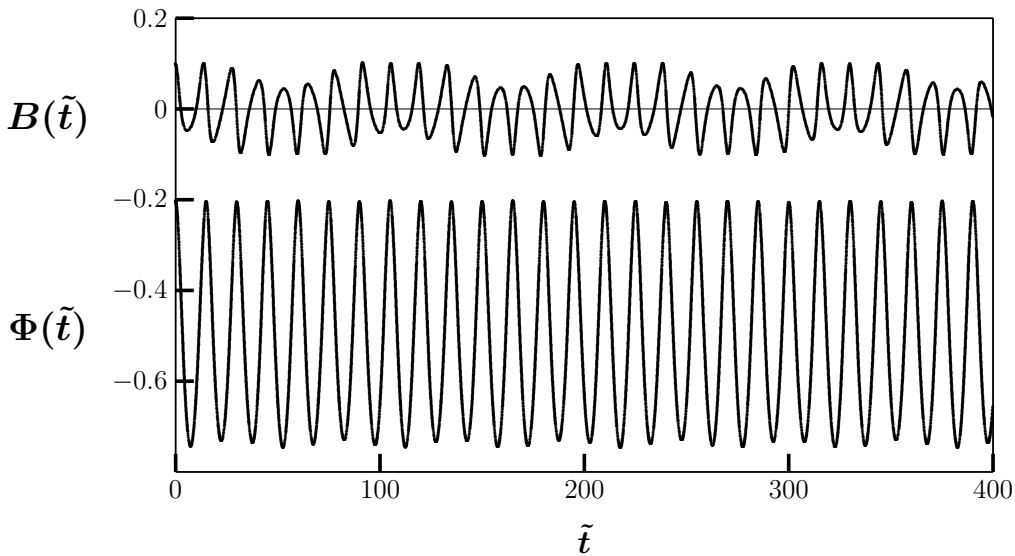}
\caption{$\Phi$ and $B$ in the Einstein frame for $ \lambda
=-1/2$, $B_0=0.1$, and  $\Phi_0=-0.2.$ The dilaton stabilization
is observed as well as the periodically waving oscillation of
$B$.} \label{fig3}
\end{figure}

When $\lambda$ is negative, there are two classes of solutions,
depending on the initial value of the dilaton $\Phi_0$
(or equivalently the energy density) for a fixed initial magnetic
condensation $B_0$. For large  $\Phi_0$ or $B_0$, the
configurations are almost the same as those for nonnegative
$\lambda$ (see Figure~\ref{fig2}).
For small enough $\Phi_0$ and $B_0$,
the dilaton shows the
oscillating behavior as expected from the
dilaton potential in Figure~\ref{fig1}. According to the dilaton
oscillation, the magnetic component is also oscillating around a
condensed value (see Figure~\ref{fig3}). If we involve the radiation
of perturbative modes of the dilaton and the antisymmetric tensor
field, which is missing in this classical configuration, the
classical solution will damp to a stabilized value of the dilaton
and the final condensed value of the magnetic component.

\section{Cosmological Homogeneous Solutions}
\label{sec:4}

In this section we study cosmological solutions of the D-brane
universe in the presence of both the dilaton and the antisymmetric
tensor field. As we have done in the previous paper
\cite{Chun:2005ee}, we assume spatially homogeneous configurations
for the antisymmetric tensor field and the dilaton, and look for the
time evolution of these fields and the expansion of the universe.

The non-vanishing homogeneous antisymmetric tensor field, in
general, implies the anisotropic universe. To consider the simplest
form of anisotropic cosmology, we take only a single magnetic component of
${\cal B}_{\mu\nu}$ to be nonzero, namely ${\cal B}_{12}(t)\equiv B(t)$
and ${\cal B}_{0i}(t)={\cal B}_{23}(t)={\cal B}_{31}(t)=0$. Then the metric
consistent with this choice of field configuration is of Bianchi
type I
\begin{equation}
ds^2 = -dt^2 + a_1(t)^2(dx^1)^2 + a_2(t)^2(dx^2)^2 +
a_3(t)^2(dx^3)^2.
\end{equation}
Then, in the Einstein frame, the field equation for $B$ (\ref{B-eq1}) reads
\begin{equation}
\frac{d^{2}B}{dt^{2}} + \left(-\frac{1}{a_{1}}\frac{da_{1}}{dt}
-\frac{1}{a_{2}}\frac{da_{2}}{dt}
+\frac{1}{a_{3}}\frac{da_{3}}{dt}-4\frac{d\Phi}{dt}\right)
\frac{dB}{dt}
+\frac{m_B^2e^{3\Phi}B}{\sqrt{1+B^2/e^{4\Phi}a_1^2a_2^2}} = 0,
\label{eq=Beq}
\end{equation}
the dilaton-field equation (\ref{P-eq1}) is
\begin{equation}
\frac{d^{2}\Phi}{dt^{2}}+\left(
\frac{1}{a_{1}}\frac{da_{1}}{dt}
+\frac{1}{a_{2}}\frac{da_{2}}{dt}
+\frac{1}{a_{3}}\frac{da_{3}}{dt}\right)
\frac{d\Phi}{dt} = -2\varrho_B  -
\frac12\varrho_b - \tilde\varrho_b- \varrho_\Lambda, \label{eq=Phieq}
\end{equation}
and Einstein equations (\ref{E-eq1}) are
\begin{eqnarray}
\frac{1}{a_{1}}\frac{da_{1}}{dt}\frac{1}{a_{2}}\frac{da_{2}}{dt}
+\frac{1}{a_{2}}\frac{da_{2}}{dt}\frac{1}{a_{3}}\frac{da_{3}}{dt}
+\frac{1}{a_{3}}\frac{da_{3}}{dt}\frac{1}{a_{1}}\frac{da_{1}}{dt}
&=& \varrho_\Phi +\varrho_B +\varrho_b + \varrho_\Lambda, \label{eq=G00}\\
\frac{1}{a_{2}}\frac{d^{2}a_{2}}{dt^{2}}
+\frac{1}{a_{3}}\frac{d^{2}a_{3}}{dt^{2}}
+\frac{1}{a_{2}}\frac{da_{2}}{dt}\frac{1}{a_{3}}\frac{da_{3}}{dt}
&=& -\varrho_\Phi +\varrho_B +\tilde{\varrho}_b + \varrho_\Lambda,
\label{eq=G11}\\
\frac{1}{a_{3}}\frac{d^{2}a_{3}}{dt^{2}}
+\frac{1}{a_{1}}\frac{d^{2}a_{1}}{dt^{2}}
+\frac{1}{a_{3}}\frac{da_{3}}{dt}\frac{1}{a_{1}}\frac{da_{1}}{dt}
&=& -\varrho_\Phi +\varrho_B +\tilde{\varrho}_b + \varrho_\Lambda,
\label{eq=G22}\\
\frac{1}{a_{1}}\frac{d^{2}a_{1}}{dt^{2}}
+\frac{1}{a_{2}}\frac{d^{2}a_{2}}{dt^{2}}
+\frac{1}{a_{1}}\frac{da_{1}}{dt}\frac{1}{a_{2}}\frac{da_{2}}{dt}
&=& -\varrho_\Phi -\varrho_B +\varrho_b + \varrho_\Lambda, \label{eq=G33} %
\end{eqnarray}
where
\begin{eqnarray}
&\displaystyle
\varrho_\Phi = \left(\frac{d\Phi}{dt}\right)^2\,,\quad %
\varrho_B = \frac{e^{-4\Phi}}{4a_1^2a_2^2} \left(\frac{dB}{dt}\right)^{2}\,,
\quad %
\varrho_\Lambda = \Lambda e^{2\Phi}\,,
\label{br1} &\\&\displaystyle
\varrho_b = \frac12m_B^2e^{3\Phi}\left(1+\frac{e^{-4\Phi}B^2}{a_1^2a_2^2}\right)^{1/2}\,,\quad %
\tilde\varrho_b = \frac12m_B^2e^{3\Phi}
\left(1+\frac{e^{-4\Phi}B^2}{a_1^2a_2^2}\right)^{-1/2}\,. &%
\label{br2}
\end{eqnarray}
It is more convenient to employ the dimensionless time variable $\tilde t$
of (\ref{ti}) and to introduce the variables $\alpha_i$ and $b$ defined by
\begin{equation}
\alpha_i=\ln a_i,\qquad b=\frac{e^{-2\Phi}B}{a_1a_2}.
\end{equation}
Then the above equations (\ref{eq=Beq})--(\ref{eq=G33}) are rewritten as
\begin{eqnarray}
\ddot b + (\dot\alpha_1+\dot\alpha_2+\dot\alpha_3)\dot b + \Bigg[
2\ddot\Phi+2(-\dot\alpha_1-\dot\alpha_2+\dot\alpha_3-2\dot\Phi)\dot\Phi
\hspace{15mm}&&\nonumber\\ \label{eq-b}
+\ddot\alpha_1+\ddot\alpha_2+(\dot\alpha_1+\dot\alpha_2)\dot\alpha_3
+\frac{e^{3\Phi}}{\sqrt{1+b^2}}\Bigg]b &=& 0,
\end{eqnarray}
\begin{equation}
\label{eq-phi}%
\ddot\Phi + (\dot\alpha_1+\dot\alpha_2+\dot\alpha_3)\dot\Phi =
-2\rho_B-\rho_\Lambda-\frac12\rho_b-\tilde\rho_b,
\end{equation}
\begin{eqnarray}
\dot\alpha_1\dot\alpha_2+\dot\alpha_2\dot\alpha_3+\dot\alpha_3\dot\alpha_1
&=& \rho_\Phi+\rho_B+\rho_\Lambda+\rho_b, \label{eq-constraint}\\
\ddot\alpha_1+\dot\alpha_1(\dot\alpha_1+\dot\alpha_2+\dot\alpha_3)
&=& \rho_\Lambda+\rho_b, \\
\ddot\alpha_2+\dot\alpha_2(\dot\alpha_1+\dot\alpha_2+\dot\alpha_3)
&=& \rho_\Lambda+\rho_b, \\
\ddot\alpha_3+\dot\alpha_3(\dot\alpha_1+\dot\alpha_2+\dot\alpha_3)
&=& 2\rho_B+\rho_\Lambda+\tilde\rho_b, \label{eq-a3}
\end{eqnarray}
where, by dividing $m_{B}^{2}$ from each $\varrho$ in
Eqs.~(\ref{br1})--(\ref{br2}), we have
\begin{eqnarray}
&\displaystyle
\rho_\Phi=\dot\Phi^2,\quad%
\rho_B=\frac14\left[\dot
b+(\dot\alpha_1+\dot\alpha_2+2\dot\Phi)b\right]^2,\quad
\rho_\Lambda=\lambda e^{2\Phi}, &\nonumber\\&\displaystyle
\rho_b=\frac12e^{3\Phi}(1+b^2)^{1/2},\quad
\tilde\rho_b=\frac12e^{3\Phi}(1+b^2)^{-1/2}, &
\end{eqnarray}
where $\lambda$ is defined in (\ref{co}).
In the subsequent subsections, we look for the solutions to the above
field equations for various cases beginning with some simple
solutions.

\subsection{Attractor solutions}
\label{subsec:4.A}

The dilaton in the system of equations (\ref{eq-b})--(\ref{eq-a3})
has the exponential potential up to the correction due to the
antisymmetric tensor field. It is well-known that the scalar field
with the exponential potential possesses the scaling solution in
which the energy density of the scalar field mimics the background
fluid energy density \cite{Copeland:1997et,Copeland:2006wr}. This scaling solution is
also an attractor, so that the late time behavior of the
solutions are universal irrespective of initial conditions.
This is an attractive property of the exponential potential. For the
potential of the form $V(\Phi)=V_0e^{\beta\Phi}$, there is an
attractor solution
\begin{equation}
\label{sol-a}
\Phi=-\frac{2}{\beta}\ln\left[\frac{\beta
V_0^{1/2}t}{\sqrt{2(12-\beta^2)}}\right],\quad
\alpha=\left(\frac{2}{\beta}\right)^2\ln t,
\end{equation}
for $0\le\beta<\sqrt{12}\,$. The scale factor obeys the power-law
time-dependence, implying that the rolling of $\Phi$ constitutes
the matter having an equation of state $p=w\rho$ where $w=\beta^{2}/6-1$
varies from $-1$ to $+1$ for the aforementioned range of $\beta$.

We found this type of particular solutions
of the Eqs.~(\ref{eq-b})--(\ref{eq-a3}),
which can be found when we have a single exponential term in the potential,
that is, for the case of $\Lambda=0$ and for the case of $m_B=0$.
For both cases we start from an ansatz of the form
\begin{equation}
\Phi(\tilde t)=\gamma_\Phi\ln\tilde t+\Phi_0,\quad%
\alpha_1=\gamma_1\ln\tilde t,\quad%
\alpha_3=\gamma_3\ln\tilde t,\quad%
b={\rm constant},
\end{equation}
where we suppressed constant terms in $\alpha_1$ and $\alpha_3$
which correspond to simple rescaling of coordinates. For the case of
$\Lambda=0$, we obtain two distinguished solutions
\begin{equation}
\label{sol-1}
\Phi(\tilde t)=-\frac{2}{3}\ln\tilde t+\ln\frac{2}{3},\quad%
\alpha_1(\tilde t)=\alpha_3(\tilde t)=\frac{4}{9}\ln\tilde t,\quad%
b=0,
\end{equation}
and
\begin{equation}
\label{sol-2}%
\Phi(\tilde t)=-\frac{2}{3}\ln\tilde t
 + \frac{2}{3}\ln\left(\frac{8\cdot5^{1/4}}{21}\right),\quad%
\alpha_1(\tilde t)=\frac{10}{21}\ln\tilde t,\quad%
\alpha_3(\tilde t)=\frac{3}{7}\ln\tilde t,\quad%
b=\pm\frac12.
\end{equation}
The first solution is nothing but the solution (\ref{sol-a}) with $\beta=3$.
The second solution has the non-vanishing antisymmetric tensor field.
For the case of $m_B=0$, we introduce a new rescaled dimensionless time
variable ${\bar t}=\Lambda^{1/2}t$ instead of ${\tilde t}$ of (\ref{ti})
and then get the continuous set of solutions
from Eqs.~(\ref{eq=Beq})--(\ref{eq=G33})
\begin{equation}
\label{sol-3}%
\Phi({\bar t})=-\ln{\bar t}+\frac{1}{2}\ln2,\quad%
\alpha_1({\bar t})=\alpha_3({\bar t})=\ln{\bar t},\quad%
b={\rm arbitrary\ constant}.
\end{equation}
$\Phi({\bar t})$ and $\alpha({\bar t})$ are same as those in
Eq.~(\ref{sol-a}) with $\beta=2$,
while we have the non-vanishing $B$ field condensate.

These solutions are the solutions to the
Eqs.~(\ref{eq-b})--(\ref{eq-a3}) for the specific initial
conditions. However, the importance of these solutions, as noted in
the first paragraph, arises from the fact that they are attractors,
which means that after enough time the solutions with different initial
conditions approach these solutions.
We will confirm this through numerical analysis in the next subsection.

The solution (\ref{sol-1}) applies for the brane tension dominated case
where the dilaton potential is approximated by $m_{B}^{2}U(\Phi)
=\frac12m_B^2e^{3\Phi}$ from Eq.~(\ref{pu}).
The evolution of the dilaton under this potential produces matter with
the equation of state $p=\frac12\rho$.

Once the antisymmetric tensor field is turned on, the anisotropy
appears as in the solution (\ref{sol-2}).
The measure of anisotropy is
\begin{equation}
\frac{\dot\alpha_3}{\dot\alpha_1}=\frac{9}{10}.
\end{equation}
This result is contrasted with that in Ref.~\cite{Chun:2005ee}
where the dilaton is assumed to be stabilized.
The rolling of the dilaton makes the difference.
It affects the dynamics of $B$ field in such
a way that $b$ remains constant at $b=\pm1/2$ instead of oscillating
about the potential minimum $b=0$ and the anisotropy is maintained.

Let us turn to the third solution (\ref{sol-3}). It is relevant when
the dilaton potential arising from the curvature of extra
dimensions, $V(\Phi)=\Lambda e^{2\Phi}$, dominates over other
contributions. In our scheme this happens as the dilaton rolls down
the potential. When $b$ vanishes, the transition point at which
$\Lambda e^{2\Phi}$ starts to dominate over $\frac12m_B^2e^{3\Phi}$ is at
$\Phi_t=\ln(2\lambda)$. Thus, this solution describes the late time
behavior of all the solutions with various initial conditions when
$\Lambda$ is positive.
It is very intriguing since we achieve the isotropic
universe in the end. The rolling of dilaton makes the brane tension
term which causes the anisotropy when we have non-vanishing $B$ field
less important than the extra dimension curvature term which recovers
the isotropy. The rolling of the dilaton under the potential
$\Lambda e^{2\Phi}$ now forms the matter having
the equation of state $p=-\frac13\rho$, thus giving marginal inflation.

\subsection{Numerical analysis}
\label{subsec:4.B}

\subsubsection{Initial conditions}
\label{subsubsec:4.B.1}

We have the second order differential equations for four variables
$\Phi(t)$, $B(t)$, $\alpha_1(t)=\alpha_2(t)$, $\alpha_3(t)$. Thus we
need eight initial values $\Phi_0$, $\dot\Phi_0$, $B_0$, $\dot B_0$,
$\alpha_{10}$, $\dot\alpha_{10}$, $\alpha_{30}$, $\dot\alpha_{30}$
to specify the solution. Among these, $\alpha_{i0}$ can always be
set to zero by coordinate rescaling. $\dot\alpha_{i0}$ must obey
the constraint equation (\ref{eq-constraint}), but this does not fix
the ratio $\dot\alpha_{10}/\dot\alpha_{30}$.
We choose the isotropic universe as a natural initial condition
which leads to $\dot\alpha_{10}=\dot\alpha_{30}$.

Since the dilaton potential is composed of exponential terms, the shift of
the dilaton field by a constant can be traded for the redefinition of
mass scale. We use this property to take the initial value of the
dilaton to be zero without loss of generality. In our numerical
analysis, we take the dimensionless time variable as
$\tilde t\equiv\tilde mt$ where $\tilde m=m_Be^{\frac32\Phi_0}$
and use the variable $\tilde\Phi\equiv\Phi-\Phi_0$ with its initial value
$\tilde\Phi_0=0$. This means that the proper time scale for the
cosmological evolution is not $m_B^{-1}$, but $\tilde m^{-1}$.
For convenience's sake, we take the initial time as $\tilde t_0=1$.
The other mass scale $\Lambda$ is also affected by this shift
and we can treat it by replacing the parameter $\lambda$ with
$\tilde\lambda\equiv\lambda e^{-\Phi_0}$.

Now we need three initial values $\dot\Phi_0$, $B_0$, and $\dot
B_0$, to fix the functional form of the solution.
The initial values $b_0$ and $\dot b_0$ are related to
$B_0$ and $\dot B_0$ by $b_0=B_0$ and
$\dot b_0=\dot B_0-2(\dot\alpha_{10}+\dot\Phi_0)B_0$.

\subsubsection{Numerical solutions for $B=0$}
\label{subsubsec:4.B.2}

If we assume ${\cal B}_{\mu\nu}=0$, we can study the role of the
dilaton interacting with gravity more clearly as was also true
in the subsection~\ref{subsec:3.A}. We keep the brane
term with the tension $m_B$. The spacetime is now isotropic. The
parameter $\Lambda$ in the string-frame action, which can be
interpreted as the curvature of the compact internal manifold, plays
a very important role, and the solutions are topologically different
depending on the signature of $\Lambda$.

\begin{figure}
\begin{center}
\includegraphics[height=65mm]{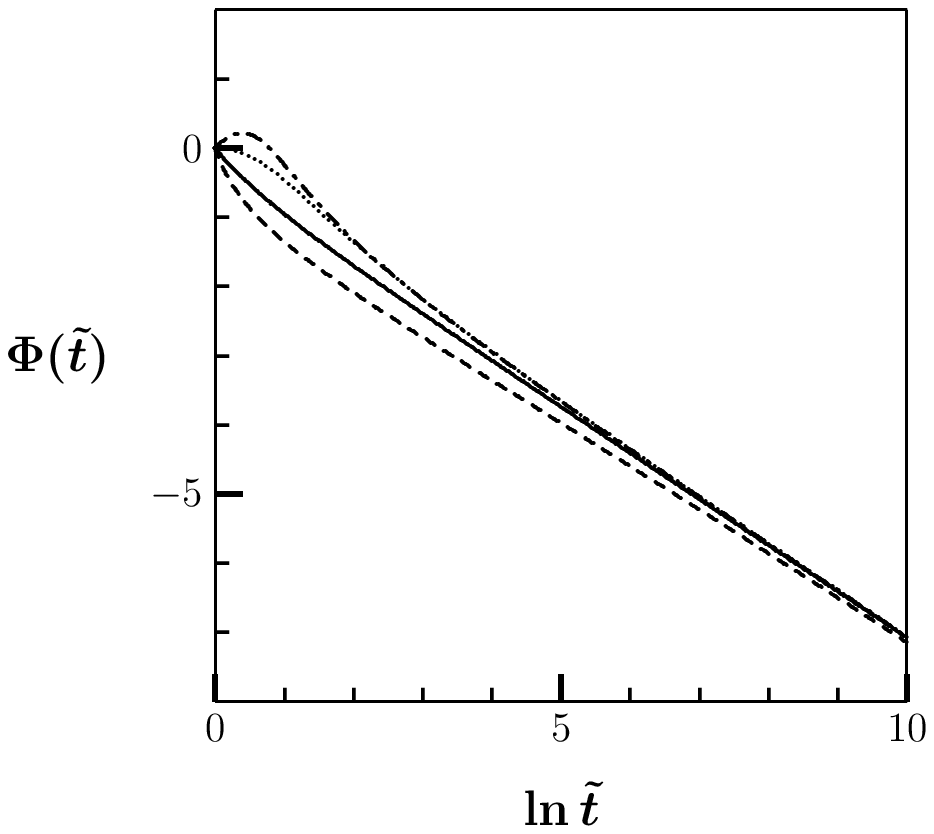}\hspace{5mm}
\includegraphics[height=65mm]{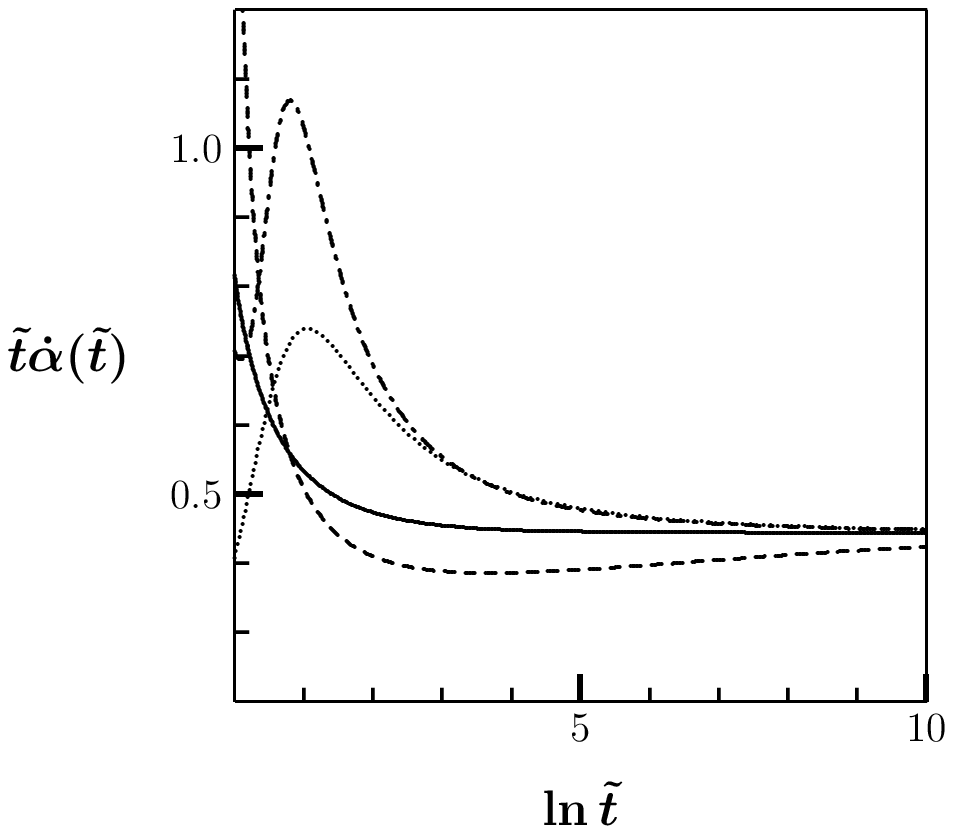}
\end{center}
\caption{Numerical solutions for $B=0$ and $\Lambda=0$ with initial
value $\dot\Phi_0=-\sqrt{3/2}$ (solid curve, the attractor), $0$
(dotted curve), $-3$ (dashed curve) and $1$ (dash-dotted curve),
respectively. For the evolution of the scale factor, we plot
$\tilde t \dot\alpha(\tilde t)$ for it becomes constant $\gamma$
when the scale factor behaves as $a\propto t^\gamma$.}
\label{fig4}
\end{figure}

When $\Lambda=0$, we have an attractor solution (\ref{sol-1}). Since
we consider the dilaton only, the only relevant initial condition is
$\dot\Phi_0$. The initial condition for the attractor solution is
$\dot\Phi_0=-\sqrt{3/2}$ in the setup described in the previous subsection.
For a few other values of $\dot\Phi_0$, we plot the numerical solutions in
Figure~\ref{fig4}. Regardless of the initial value $\dot\Phi_0$,
all the solutions approach universally the attractor solution (up to
rescaling for the scale factor) after some time.

\begin{figure}
\begin{center}
\includegraphics[height=65mm]{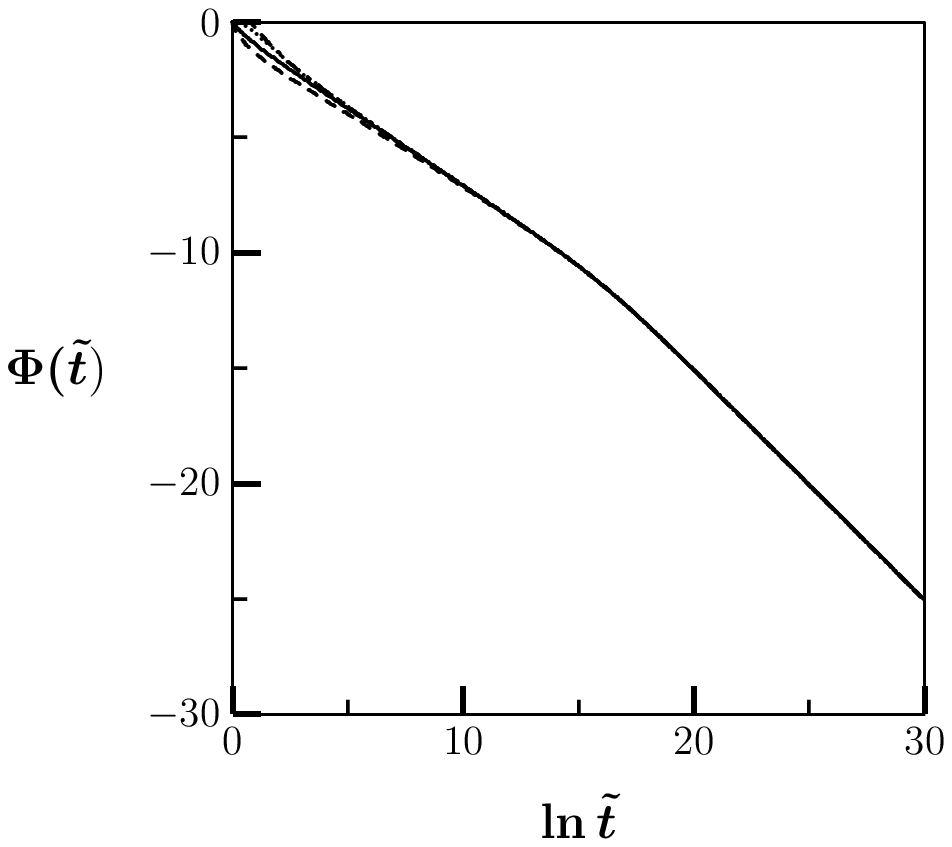}\hspace{5mm}
\includegraphics[height=65mm]{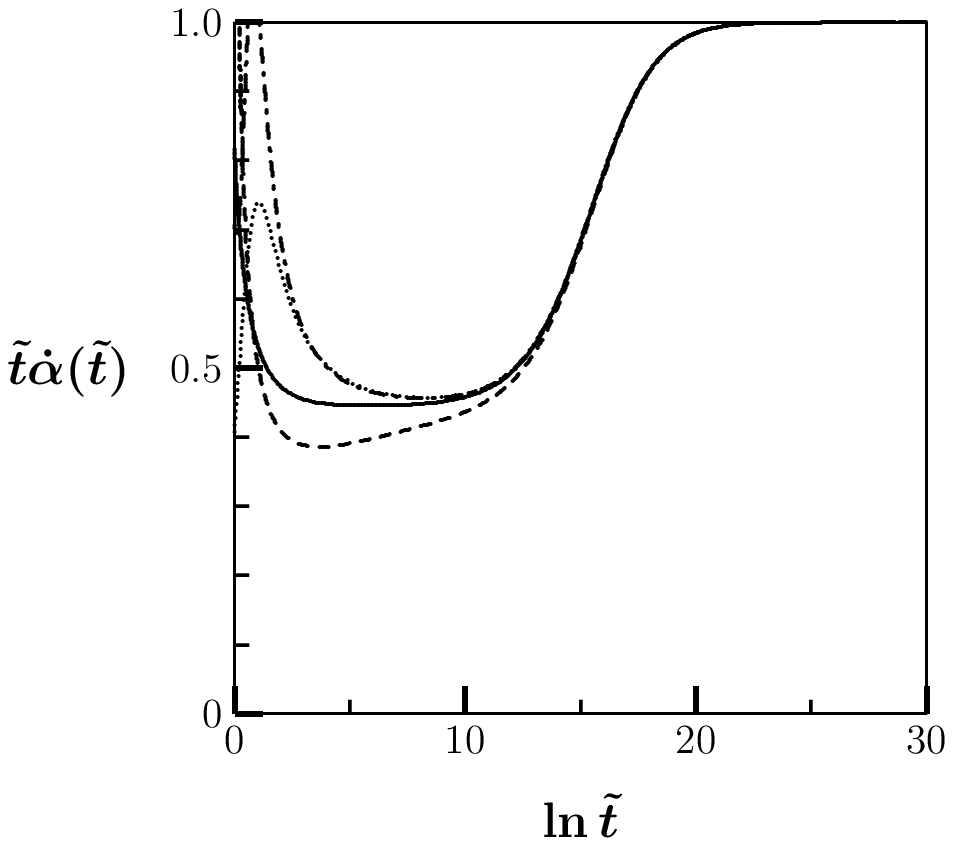}
\end{center}
\caption{Numerical solutions for $B=0$ and $\Lambda>0$.
When we set the parameter $\lambda=10^{-4}$, the curves stand for the
solutions with the initial
value $\dot\Phi_0=-\sqrt{3/2}$ (solid curve) and $0$ (dotted curve),
$-3$ (dashed curve) and $1$ (dash-dotted curve), respectively.}
\label{fig5}
\end{figure}

When $\Lambda>0$, the evolution is divided into two stages.
When $\Phi$ is large so that $\rho_b=\frac12e^{3\Phi}$ is much larger
than $\rho_\Lambda=\lambda e^{2\Phi}$,
the solution approaches the attractor (\ref{sol-1}).
As $\Phi$ rolls down to around $\Phi_t=\ln(2\lambda)$,
$\rho_\Lambda$ becomes larger than $\rho_b$.
After that point the solution approaches the attractor (\ref{sol-3})
with $b=0$.
In Figure~\ref{fig5},
numerical solutions show this two stage evolution explicitly.

Since $\Lambda$ measures the curvature of the extra dimensions,
its value can be negative.
When $\Lambda<0$, the dilaton potential has
the global minimum at $\Phi_{{\rm min}}=\ln(-4\lambda/3)$ with
negative cosmological constant $U_{{\rm min}}$ as discussed in the
subsection~\ref{subsec:3.A}. The disastrous consequence of this
negative energy minimum is that the scale factor collapses in the end.
This cannot be avoided because the dilaton rolls down toward the minimum
and stays where the potential is negative while the kinetic energy
is diluted by the expansion, thus the total energy becomes negative
and it derives the universe to collapse. It is different from the case of flat
spacetime in the subsection~\ref{subsec:3.A},
where the dilaton experiences either the permanent oscillation around
$\Phi_{{\rm min}}$ for negative initial energy density or
the attractor solution for nonnegative initial energy density.

\subsubsection{Numerical solutions for $B\ne0$}
\label{subsubsec:4.B.3}

Now, we turn on the antisymmetric tensor field along the $x^3$
direction, ${\cal B}_{12}=B\ne 0$. The spacetime becomes anisotropic,
$a_1(t)=a_2(t)\ne a_3(t)$, in general.
The solutions are classified again by the signature of $\Lambda$.

\begin{figure}
\begin{center}
\includegraphics[height=65mm]{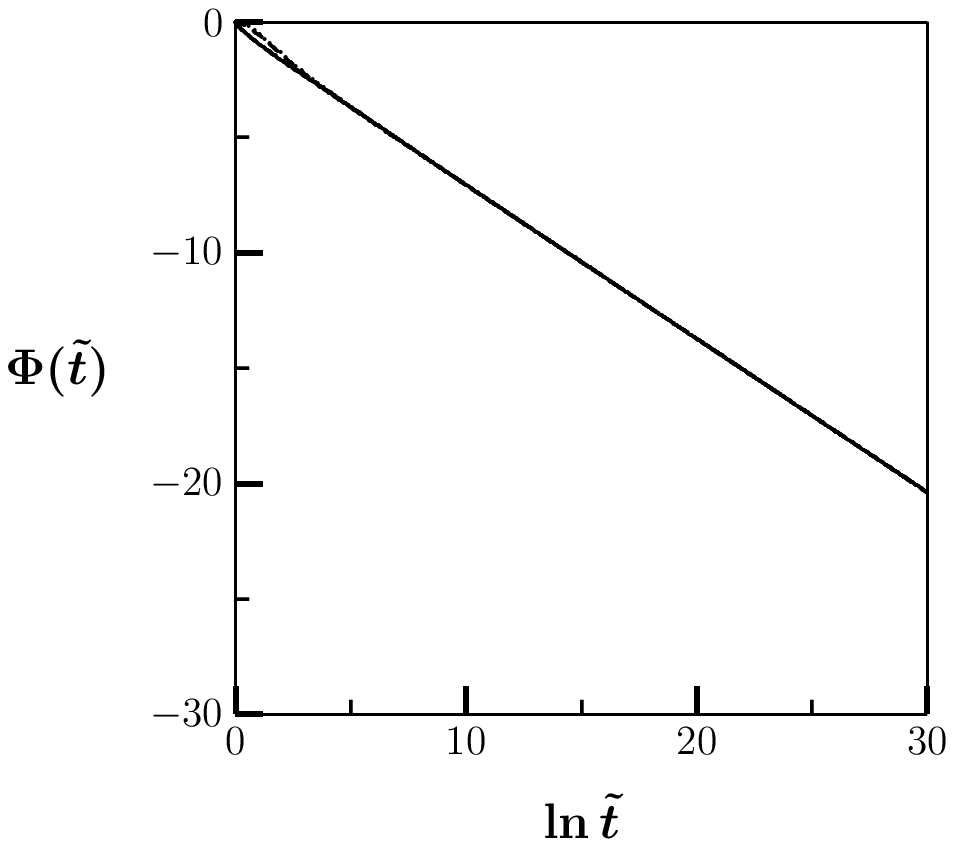}\hspace{5mm}
\includegraphics[height=65mm]{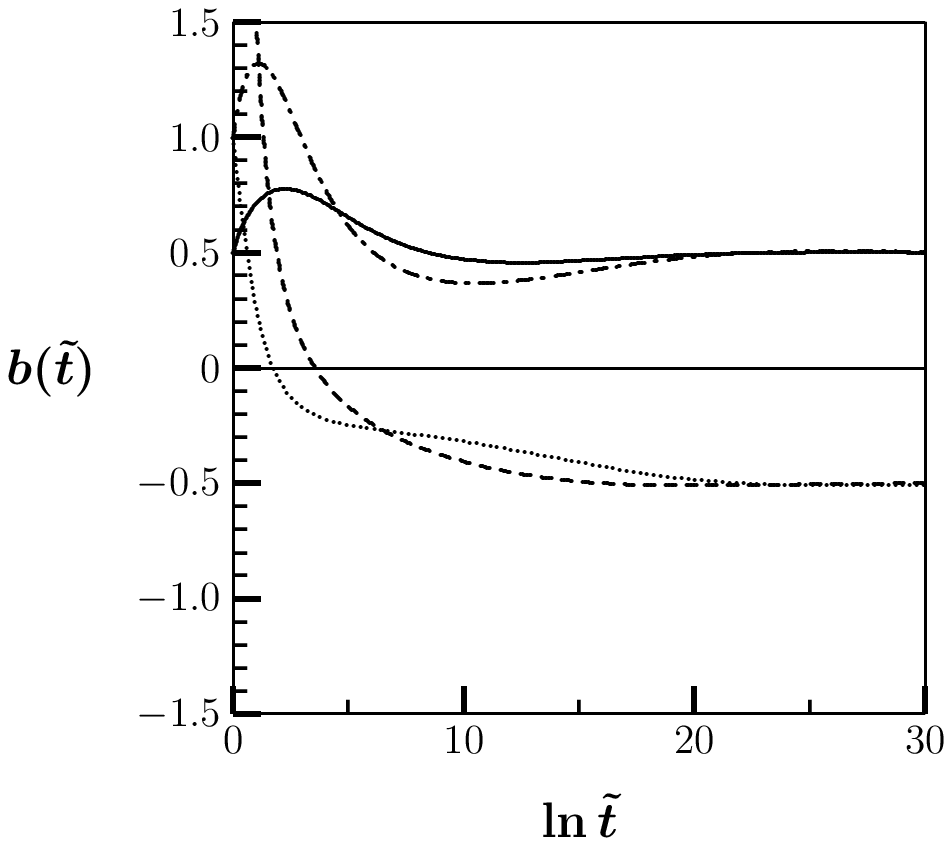}\\[10mm]
\includegraphics[height=65mm]{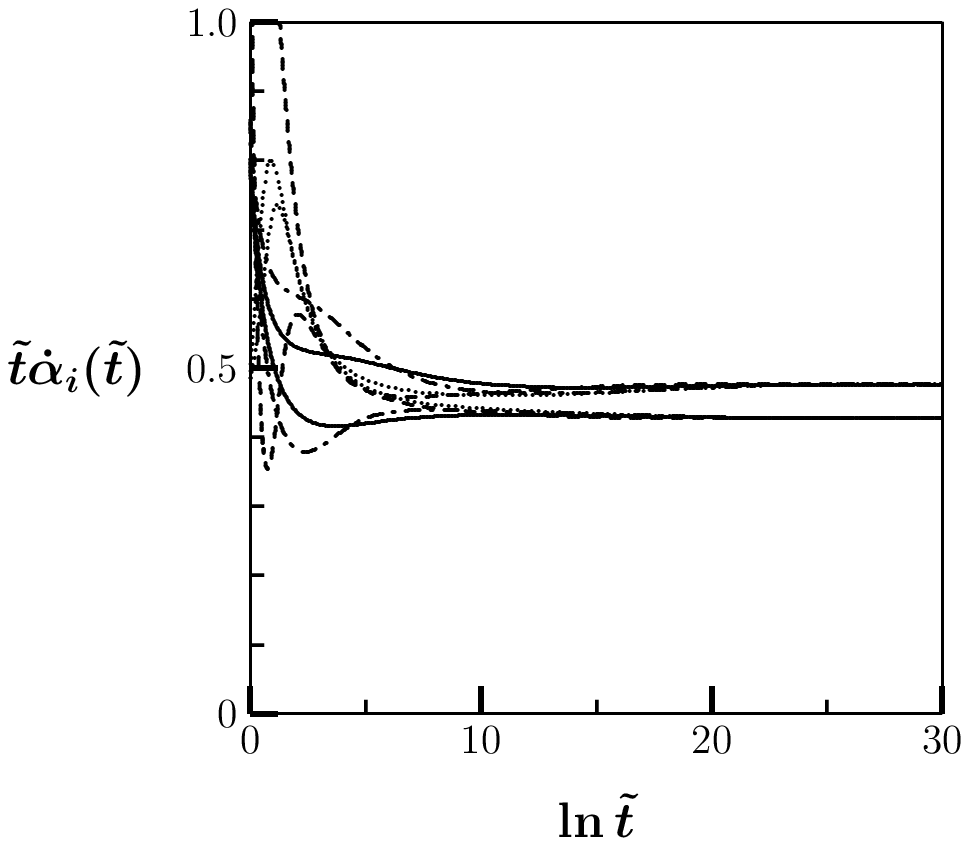}\hspace{5mm}
\includegraphics[height=65mm]{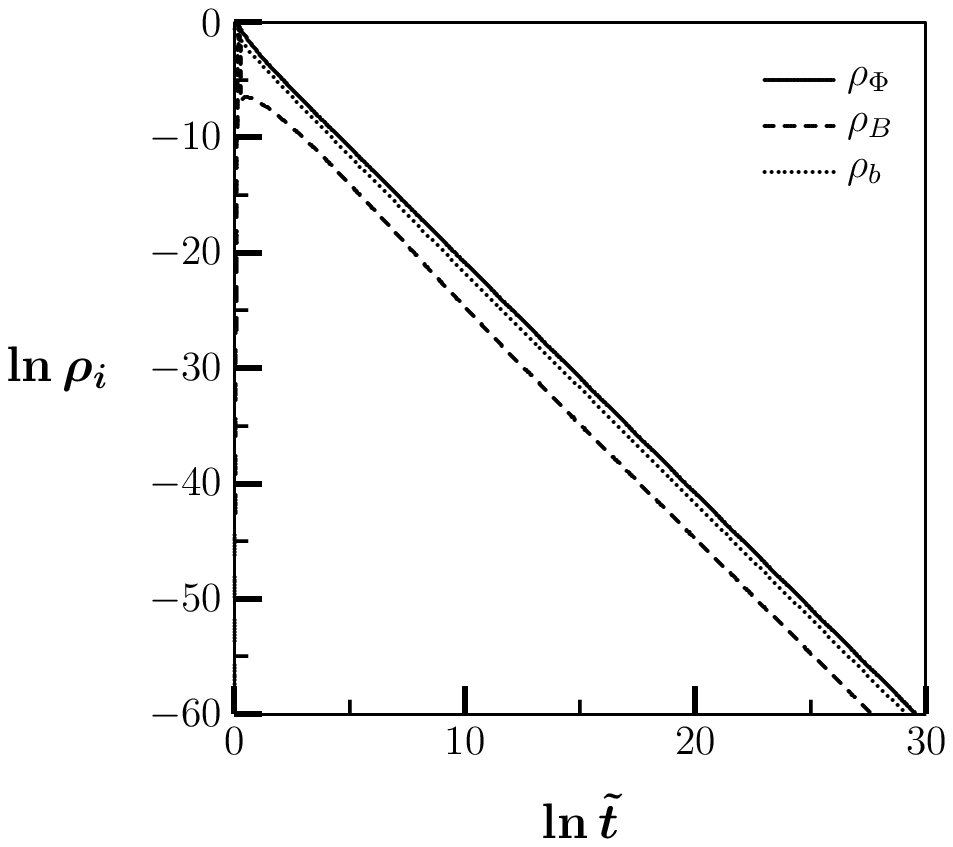}
\end{center}
\caption{Numerical solutions for $B\ne0$ and $\Lambda=0$.
The initial conditions are as follow.
Solid line: $\dot\Phi_0=-7/4\cdot5^{1/4}$, $b_0=1/2$, $\dot B_0=0$;
Dotted line: $\dot\Phi_0=0$, $b_0=1$, $\dot B_0=0$;
Dashed line: $\dot\Phi_0=0$, $b_0=10$, $\dot B_0=0$;
Dash-dotted line: $\dot\Phi_0=-\sqrt{3/2}$, $b_0=1$, $\dot B_0=0$.
The lower-right panel shows the evolution of each component of energy density
for initial conditions $\dot\Phi_0=-7/4\cdot5^{1/4}$, $b_0=1/2$, $\dot B_0=0$.}
\label{fig6}
\end{figure}

For $\Lambda=0$, we have an attractor (\ref{sol-2}) under
the initial conditions of $\dot\Phi_0=-7/4\cdot5^{1/4}$,
$b_0=\pm\frac12$, and $\dot b_0=0$.
In Figure~\ref{fig6}, we plotted numerical solutions for a few different
initial conditions in addition to the attractor.
It is confirmed again that all the solutions approach the attractor
as the time goes on.
The final value of $b$ is either $+\frac12$ or $-\frac12$
depending on the initial conditions.
In the lower-right panel in Figure~\ref{fig6} we provide the evolution of each component of
energy density. The kinetic energy of the dilaton $\rho_\Phi$ catches
up the potential energy $\rho_b$ and the ratio of them becomes constant.
This is a characteristic feature of the scaling solution \cite{Copeland:2006wr}.
The kinetic energy of $B$ field is kept much smaller than both of them,
but the anisotropy is still maintained due to the difference between $\rho_b$
and $\tilde\rho_b$.

\begin{figure}
\begin{center}
\includegraphics[height=65mm]{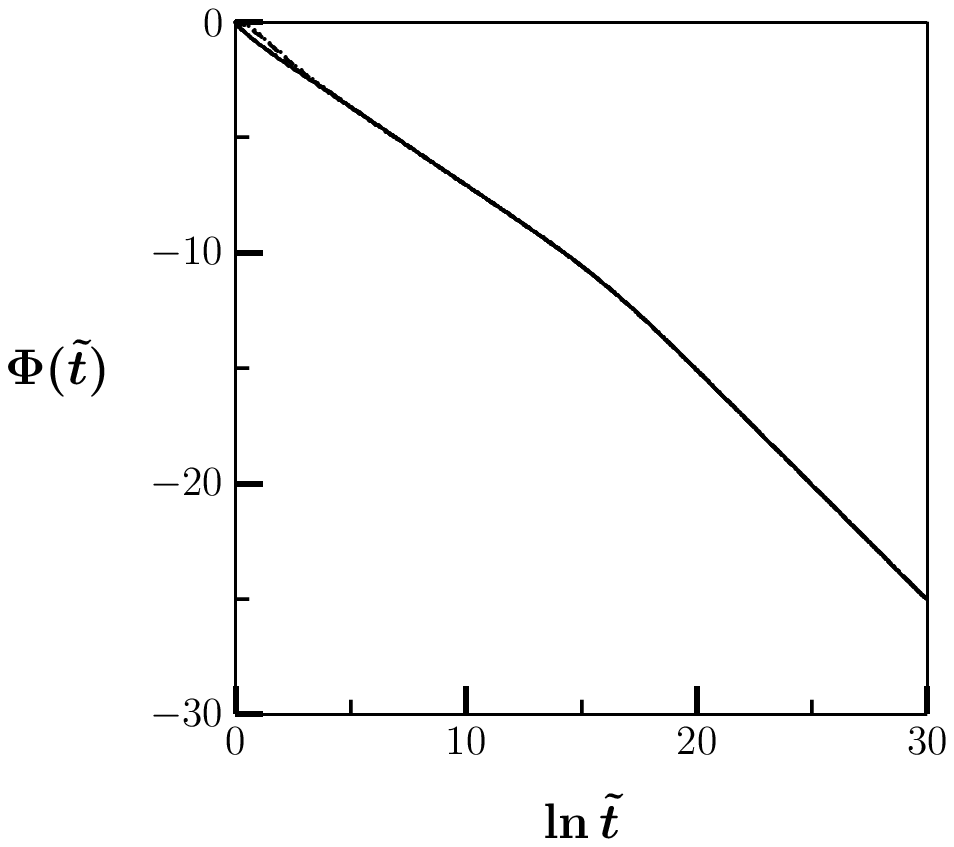}\hspace{5mm}
\includegraphics[height=65mm]{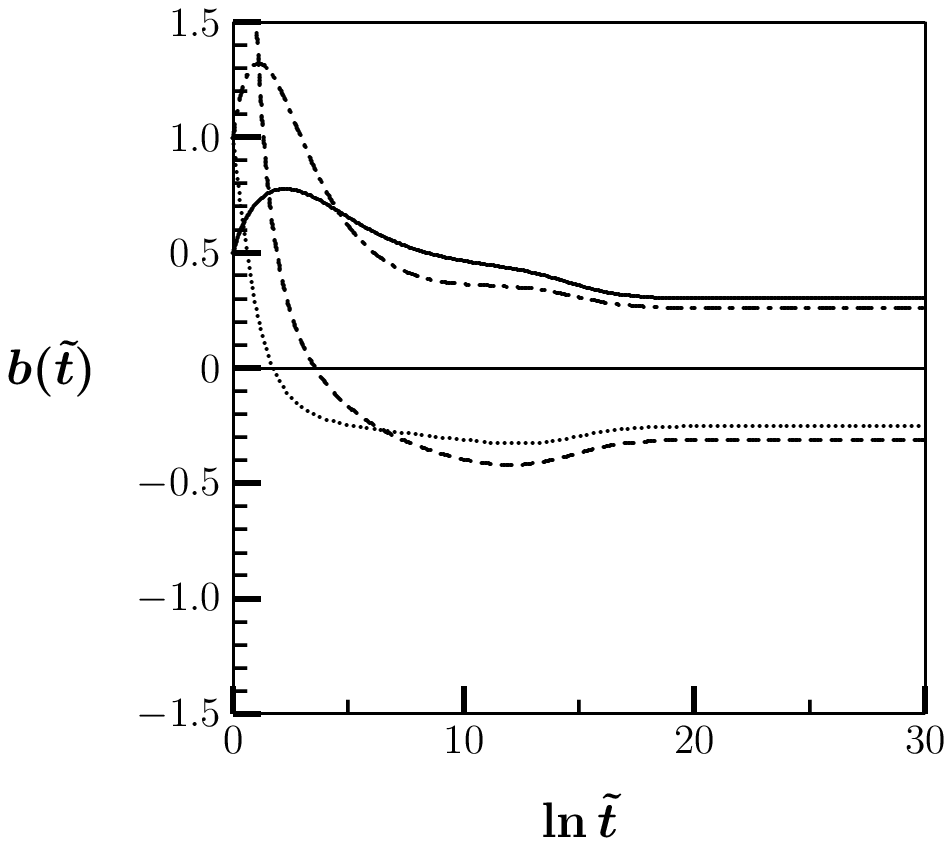}\\[10mm]
\includegraphics[height=65mm]{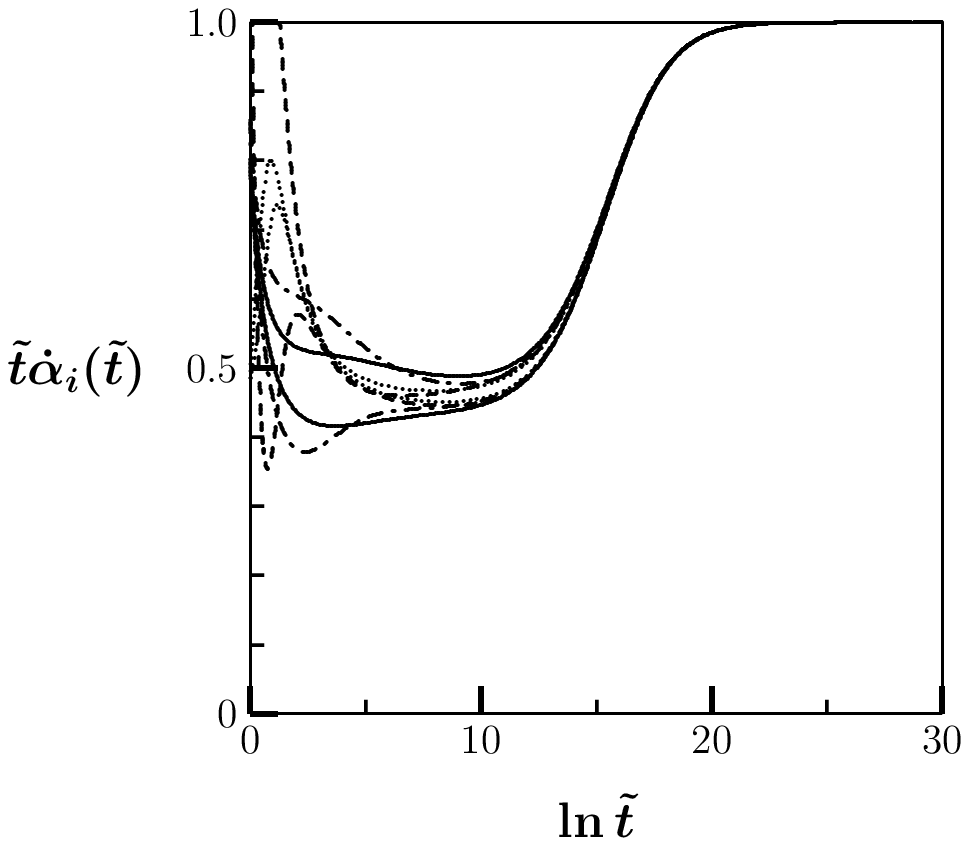}\hspace{5mm}
\includegraphics[height=65mm]{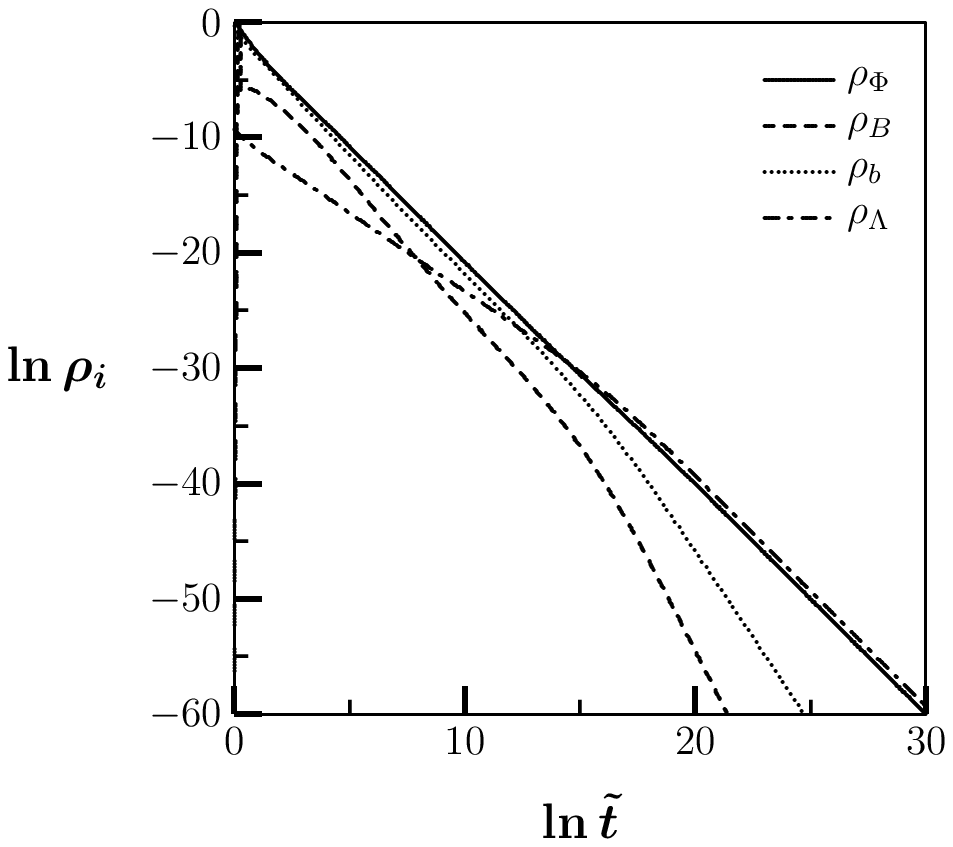}
\end{center}
\caption{Numerical solutions for $B\ne0$ and $\Lambda>0$.
When $\lambda=10^{-4}$, the solutions of four initial conditions are given:
$\dot\Phi_0=-7/4\cdot5^{1/4}$, $b_0=1/2$, $\dot B_0=0$ for the solid curves,
$\dot\Phi_0=0$, $b_0=1$, $\dot B_0=0$ for the dotted curves,
$\dot\Phi_0=0$, $b_0=10$, $\dot B_0=0$ for the dashed curves, and
$\dot\Phi_0=-\sqrt{3/2}$, $b_0=1$, $\dot B_0=0$ for the dash-dotted curves.
The lower-right panel shows the evolution of each component of energy density
for initial conditions $\dot\Phi_0=-\sqrt{3/2}$, $b_0=1$, $\dot B_0=0$.}
\label{fig7}
\end{figure}

For $\Lambda>0$, the evolution is again divided into two stages as in
the $B=0$ case. In the first stage where $\rho_b$ is dominant, the solution
approaches an attractor (\ref{sol-2}). In the second stage where
$\rho_\Lambda$ is dominant, it approaches another attractor (\ref{sol-3}).
Thus the universe recovers the isotropy.
The final value of $b$ is a certain constant which is determined by
initial conditions and can differ from $\pm\frac12$.
Numerical solutions for a few initial conditions
are shown in Figure~\ref{fig7}.
The kinetic energy of the dilaton $\rho_\Phi$ now catches
up the potential energy $\rho_b$ in the first stage
and $\rho_\Lambda$ in the second stage.
The ratios, $\rho_\Phi/\rho_b$ and $\rho_\Phi/\rho_\Lambda$, approach
constants in each stage. The kinetic energy of $B$ field
is kept much smaller as in $\Lambda=0$ case.

For $\Lambda<0$, the solution becomes singular as in the $B=0$ case.
Here we skip the description of such singular solutions which are not
suitable for the evolution of our Universe.

\section{Stabilized Dilaton}
\label{sec:5}

The vacuum expectation value of the dilaton determines both the gauge
and gravitational coupling constants of the low energy effective theory.
Therefore, the dilaton must be stabilized at some stage of the evolution
for the action (\ref{action-E}) to have something to do with the reality.
In this section, we study the cosmological evolution
when the dilaton is stabilized.
As for the correct mechanism of dilaton stabilization,
the consensus has not been made yet.
Our goal here is to illustrate an example of the dilaton stabilization
and look into the effect of it on the dynamics of $B$ field and
the cosmological evolution, since the overall features of which are
insensitive to the detailed mechanism of stabilization.

\begin{figure}
\begin{center}
\begin{picture}(500,430)(0,0)
\put(  0,240){\includegraphics[height=65mm]{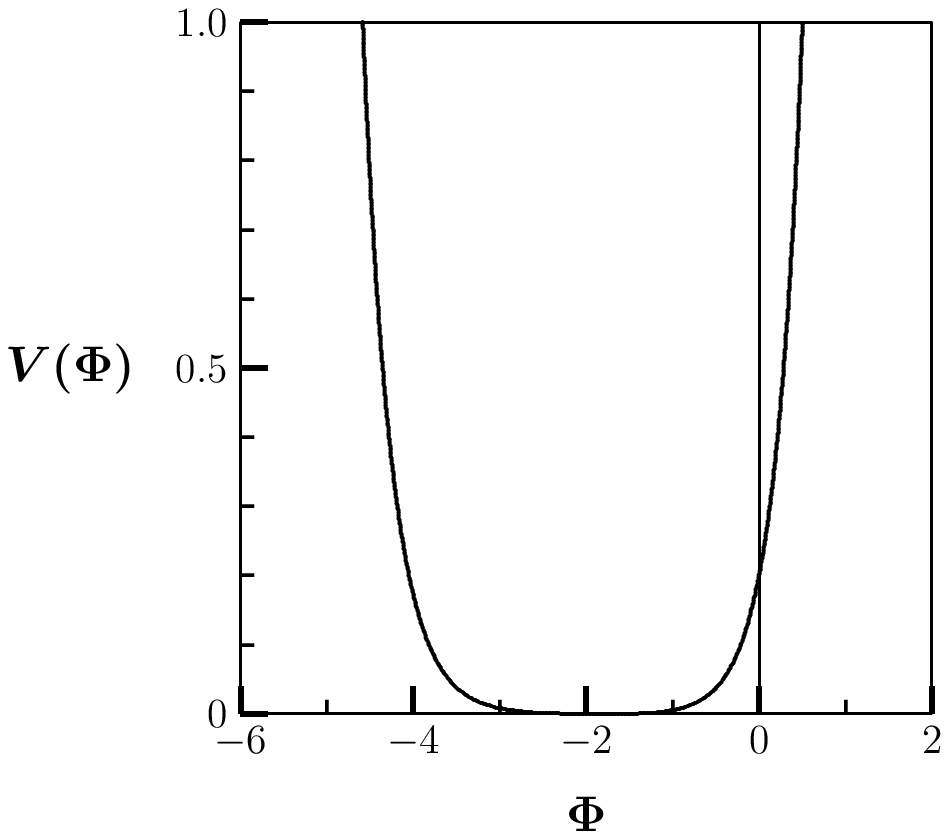}}
\put(248,240){\includegraphics[height=65mm]{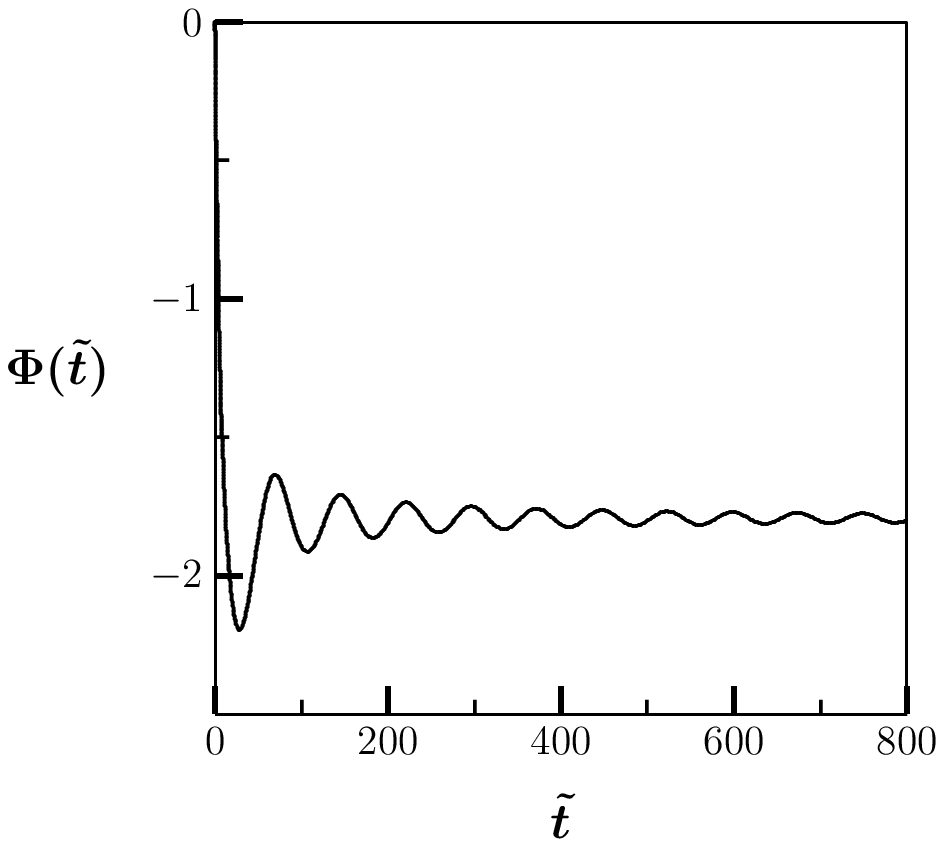}}
\put(  2,  0){\includegraphics[height=65mm]{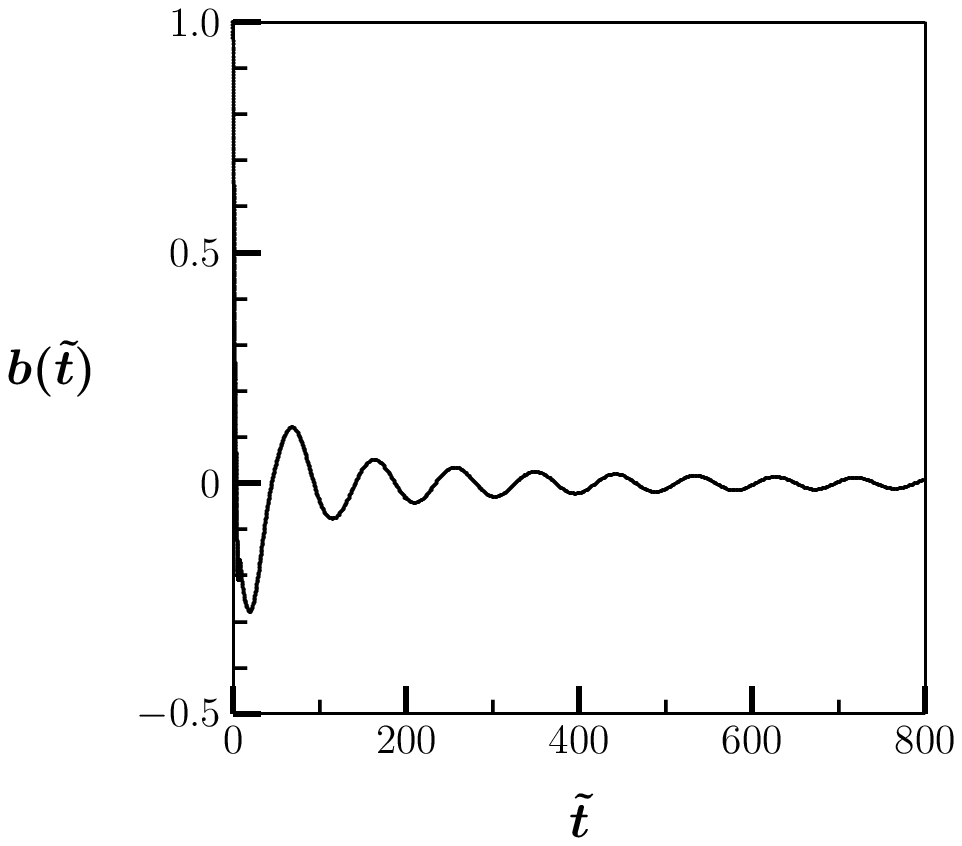}}
\put(240,  0){\includegraphics[height=65mm]{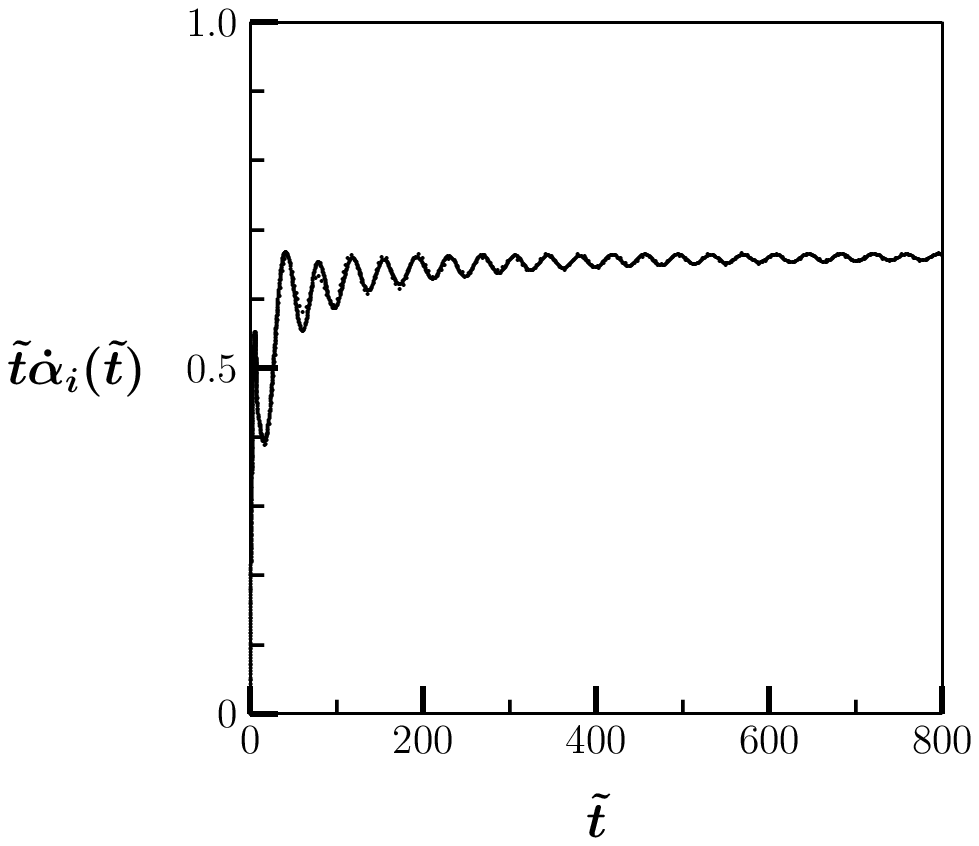}}
\put(120,435){(a)}
\put(360,435){(b)}
\put(120,195){(c)}
\put(360,195){(d)}
\end{picture}
\end{center}
\caption{(a) The potential $V(\Phi)$ for $\lambda=-0.1$.
The minimum of the potential is at $\Phi_m=-\ln 6$ with $V(\Phi_m)=0$.
(b), (c) and (d) show the numerical solutions of $\Phi(\tilde t)$,
$b(\tilde t)$ and $\dot\alpha_i(\tilde t)$, respectively,
for $\lambda=-0.1$ and initial conditions $\Phi_0=0$, $\dot\Phi_0=0$
and $b_0=1$.
$\Phi(\tilde t)$, $b(\tilde t)$, and $\tilde t\dot\alpha_i(\tilde t)$
approach $\Phi_m$, $0$, and $2/3$, respectively.}
\label{fig8}
\end{figure}

Our starting point is the dilaton potential
(\ref{dilaton-potential}). This potential possesses the minimum
for $\Lambda<0$, but the value of the potential at the minimum is
negative and need to be set to zero by fine-tuning of the constant
shift. However, the constant shift of the potential has no
motivation in the context of string theory. Instead, we introduce a term
$\frac14m_{{\rm F}}^2 e^{-3\Phi}$ in the potential, which can arise from
the effect of various form field fluxes
in extra dimensions~\cite{Lukas:1996zq}.
%\textcolor{red}{(Other important references for the potential term?)}
Thus, our potential for the dilaton for $B=0$ looks like
\begin{equation}
V_{{\rm F}}(\Phi) = \frac14\left(m_B^2e^{3\Phi}+2\Lambda e^{2\Phi}
+m_{{\rm F}}^2e^{-3\Phi}\right).
\label{eq=V1}
\end{equation}
This potential has a global minimum for any value of $\Lambda$ and
$m_{{\rm F}}$.
To have sensible cosmology, the potential at the minimum must be zero.
For $\Lambda<0$, this can be done through a fine-tuning of the parameters
in the potential
\begin{equation}
\mu^2=\frac15\left(-\frac53\lambda\right)^6,
\end{equation}
where $\mu^2=m_{{\rm F}}^2/m_B^2$.
Then the minimum is located at
\begin{equation}
\Phi_{{\rm F}}=\ln\left(-\frac53\lambda\right).
\end{equation}
The shape of this fine-tuned potential for $\lambda=-0.1$ is shown
in Figure~\ref{fig8}.
Now the equations (\ref{eq-phi})--(\ref{eq-a3}) are modified accordingly
\begin{eqnarray}
\ddot\Phi + (\dot\alpha_1+\dot\alpha_2+\dot\alpha_3)\dot\Phi
&=& -2\rho_B-\rho_\Lambda-\frac12\rho_b-\tilde\rho_b+\frac32\rho_{{\rm F}},
\label{eq-phi-mod}\\
\ddot\alpha_1+\dot\alpha_1(\dot\alpha_1+\dot\alpha_2+\dot\alpha_3)
&=& \rho_\Lambda+\rho_b+\rho_{{\rm F}}, \\
\ddot\alpha_3+\dot\alpha_3(\dot\alpha_1+\dot\alpha_2+\dot\alpha_3)
&=& 2\rho_B+\rho_\Lambda+\tilde\rho_b+\rho_{{\rm F}},
\label{eq-a3-mod}
\end{eqnarray}
where $\rho_{{\rm F}}=\frac12\mu^2e^{-3\Phi}$,
while the equation (\ref{eq-b}) for $b$ is not changed.

With the stabilizing potential for the dilaton, the cosmological evolution
is completely changed. The dilaton and the antisymmetric tensor field
rapidly come to the oscillation about the potential minimum
$\Phi=\Phi_{{\rm F}}$ and $b=0$.
Oscillating $\Phi$ and $B$ fields behave like ordinary matter satisfying
the equation of state $p=0$ \cite{Chun:2005ee}.
The universe becomes isotropic and matter dominated.
The numerical solutions for the stabilizing dilaton potential (\ref{eq=V1})
with $\lambda=-0.1$ are plotted in Figure~\ref{fig8}.
One can see the oscillation of $\Phi$ and $b$, and that both $\dot\alpha_1$
and $\dot\alpha_3$ approach to $2/3t$, indicating the matter domination.
Damping of $\Phi$ and $b$ oscillations is due to the expansion of the universe.

\section{Conclusions}
\label{sec:Conc}

We investigated  cosmology of a four-dimensional low energy
effective theory arising from the NS-NS sector of string theory
with a D-brane which contains the dynamical degrees of freedom
such as the gravity, the dilaton, and the antisymmetric tensor
field of second rank, coupling to the gauge field strength living
on the brane.
The dynamics of the system crucially depends on the curvature $\Lambda$
 and the brane tension $m_B$ through which the
dilaton obtains a potential of the form; $\Lambda e^{2\Phi} +
{1\over2} m_B^2 e^{3\Phi}$. Here, the latter becomes the effective
mass of the antisymmetric tensor field (B-matter).

In terms of the homogeneous solution in flat spacetime, we first
showed how the dilaton $\Phi(t)$ and the non-vanishing magnetic
component of the tensor field $B(t)$, in one direction evolve in
time. For positive $\Lambda$, one finds that the dilaton runs away
to negative infinity, and $B(t)$ reaches a
constant value at late time.  When $\Lambda$ is negative, one can
see that the dilaton is stabilized at a finite value and then the
dilaton (and also $B(t)$) shows an oscillatory behavior around
that value, or runs away again depending on the choice of the
initial conditions.  Such a dilaton stabilization, however,
produces a negative effective cosmological constant, and thus
leads to a collapsing universe in string cosmology.

When $B(t)$ is turned on, the universe undergoes an anisotropic expansion
described by the Bianchi type-I cosmology.
We found the attractor solutions showing the overall features of general
solutions and confirmed it through numerical analysis.
The dilaton $\Phi(t)$ decreases and settles to a logarithmic
decrease in time to negative infinity.
When the brane tension term dominates, the anisotropy is sustained.
If there is a positive curvature term, it dominates finally over the brane
tension term as the dilaton rolls down to negative infinity.
Then the expansion of the universe
turns to be isotropic and linear in time. Accordingly, $B(t)$ decreases
inversely proportional to time.

There have been various proposals to stabilize the dilaton. In order
to study the dynamics of the B-matter and the stabilized dilaton system,
we adopted an option of generating a dilaton
mass term of the form $m_{{\rm F}}^2 e^{-3\Phi}$.
Then the potential for the dilaton has a global minimum
and the cosmological constant of our Universe can
be fine-tuned to a desired value with negative $\Lambda$.
While the dilaton evolved to a
stabilized value, $B$ shows an oscillatory
matter-like behavior, and the universe expands as in the usual
matter-dominated era recovering the isotropy.
The obtained result is consistent with that of Ref.~\cite{Chun:2005ee}.

\section*{Acknowledgements}
This work is the result of research activities
(Astrophysical Research Center for the Structure and
Evolution of the Cosmos (ARCSEC)) and was supported by grant
No. R01-2006-000-10965-0 from the Basic Research Program
of the Korea Science $\&$ Engineering Foundation (I.C. $\&$ Y.K.),
by the Science Research Center Program of
the Korea Science and Engineering Foundation through
the Center for Quantum Spacetime(CQUeST) of
Sogang University with grant number R11--2005--021 (H.B.K).

\end{document}